\documentclass[runningheads,oribibl,11pt]{llncs}
\usepackage{latexsym} 
\usepackage{amsmath} 
\usepackage{amssymb} 
\usepackage{graphicx} 
\usepackage{xspace}
\usepackage{path}
\usepackage[font=small, labelfont=bf]{caption}
\usepackage[subrefformat=parens, labelfont=default]{subfig}
\usepackage{hyperref}
\usepackage{wrapfig}
\usepackage{fullpage}

\title{Polylogarithmic Approximation \\
  for Generalized Minimum Manhattan Networks} 

\author{%
  Aparna~Das\inst{1}
  \and Krzysztof~Fleszar\inst{2}
  \and Stephen~Kobourov\inst{1}
  \and Joachim~Spoerhase\inst{2}
  \and Sankar~Veeramoni\inst{1}
  \and Alexander~Wolff\inst{2}
}

\institute{
  Department of Computer Science, University of Arizona, Tucson, AZ, U.S.A.
  \and
  Lehrstuhl I, Institut f\"ur Informatik, Universit\"at W\"urzburg, Germany
}

\authorrunning{A.~Das et al.}
\titlerunning{Generalized Minimum Manhattan Networks}

\graphicspath{{pic/}}

\newcommand{\Sleft}{\ensuremath{S^\mathrm{-}}\xspace}
\newcommand{\Sright}{\ensuremath{S^\mathrm{+}}\xspace}

\newcommand{\Bleft}{\ensuremath{R^\mathrm{-}}\xspace}
\newcommand{\Bright}{\ensuremath{R^\mathrm{+}}\xspace}

\newcommand{\Nright}{\ensuremath{N^\mathrm{+}}\xspace}
\newcommand{\Nlefthor}{\ensuremath{N^\mathrm{-}_\mathrm{hor}}\xspace}
\newcommand{\Nrighthor}{\ensuremath{N^\mathrm{+}_\mathrm{hor}}\xspace}
\newcommand{\Nopt}{\ensuremath{N_\mathrm{opt}}\xspace}
\newcommand{\I}{\ensuremath{\mathcal{I}}}
\newcommand{\rsaUp}{\ensuremath{A_\mathrm{up}}\xspace}
\newcommand{\rsaDown}{\ensuremath{A_\mathrm{down}}\xspace}
\newcommand{\opt}{\mathrm{OPT}} 
\newcommand{\opthor}{\ensuremath{\opt_\mathrm{hor}}\xspace}
\newcommand{\optver}{\ensuremath{\opt_\mathrm{ver}}\xspace}

\newcommand{\topp}{\ensuremath{\mathrm{top}(I)}\xspace}
\newcommand{\bott}{\ensuremath{\mathrm{bot}(I)}\xspace}
\newcommand{\xm}{\ensuremath{m_{x}}\xspace}

\newcommand{\Rl}{\ensuremath{R_\mathrm{left}}\xspace}
\newcommand{\Rr}{\ensuremath{R_\mathrm{right}}\xspace}
\newcommand{\Al}{\ensuremath{A_\mathrm{left}}\xspace}
\newcommand{\Ar}{\ensuremath{A_\mathrm{right}}\xspace}
\newcommand{\Nl}{\ensuremath{N_\mathrm{left}}\xspace}
\newcommand{\Nr}{\ensuremath{N_\mathrm{right}}\xspace}
\newcommand{\Rm}{\ensuremath{R_\mathrm{mid}}\xspace}

\newcommand{\eps}{\ensuremath{\varepsilon}\xspace}
\newcommand{\R}{\ensuremath{\mathbb{R}}}
\newenvironment{pf}{\begin{proof}}{\qed\end{proof}}

\newtheorem{observation}{Observation}

\begin{document} 

\maketitle 

\begin{abstract}
  Given a set of $n$ terminals, which are points in $d$-dimensional
  Euclidean space, the \emph{minimum Manhattan network problem} (MMN)
  asks for a minimum-length rectilinear network that connects each
  pair of terminals by a Manhattan path, that is, a path consisting of
  axis-parallel segments whose total length equals the pair's
  Manhattan distance.  Even for $d=2$, the problem is NP-hard, but
  constant-factor approximations are known.  For $d \ge 3$, the
  problem is APX-hard; it is known to admit, for any $\eps > 0$, an
  $O(n^\eps)$-approximation.

  In the \emph{generalized minimum Manhattan network problem} (GMMN),
  we are given a set $R$ of $n$ terminal \emph{pairs}, and the goal is
  to find a minimum-length rectilinear network such that each pair in
  $R$ is connected by a Manhattan path.  GMMN is a generalization of
  both MMN and the well-known rectilinear Steiner arborescence problem
  (RSA).  So far, only special cases of GMMN have been considered.

  We present an $O(\log^{d+1} n)$-approximation algorithm for GMMN
  (and, hence, MMN) in $d \ge 2$ dimensions and an $O(\log
  n)$-approximation algorithm for 2D.  We show that an existing
  $O(\log n)$-approximation algorithm for RSA in 2D
  generalizes easily to $d>2$ dimensions.
\end{abstract}

\section{Introduction} 

Given a set of terminals, which are points in $\R^d$, the \emph{minimum
  Manhattan network problem} (MMN) asks for a minimum-length
rectilinear network that connects every pair of terminals by a
Manhattan path (\emph{M-path}, for short), that is, a path consisting
of axis-parallel segments whose total length equals the pair's
Manhattan distance. 

\begin{wrapfigure}{R}{.36\textwidth}
  \vspace*{-5ex}
  
  \hspace{1ex}\subfloat[%
  an MMN for $\{a,b,c,d,e,f\}$]%
  {\includegraphics[page=1]{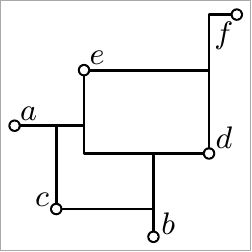}}
  \hfill
  \subfloat[%
  a GMMN for $\{(a,\!b),(c,\!d),(e,\!f)\}$]%
  {\includegraphics[page=2]{mmn-vs-gmmn}}
  \caption{MMN versus GMMN in 2D.}
  \label{fig:mmn-vs-gmmn}
  \vspace*{-3ex}
\end{wrapfigure}
In the \emph{generalized minimum Manhattan network problem} (GMMN), we
are given a set $R$ of $n$ unordered terminal \emph{pairs}, and the
goal is to find a minimum-length rectilinear network such that every
pair in~$R$ is \emph{M-connected}, that is, connected by an M-path.
GMMN is a generalization of MMN since $R$ may contain all possible
pairs of terminals.  Figure~\ref{fig:mmn-vs-gmmn} depicts examples of
both network types.

We remark that in this paper we define $n$ to be the number of
terminal \emph{pairs} of a GMMN instance, whereas previous works on
MMN defined~$n$ to be the number of \emph{terminals}.

Two-dimensional MMN (2D-MMN) naturally arises in VLSI circuit layout
\cite{gln-ammn-01}, where a set of terminals (such as gates or
transistors) needs to be interconnected by rectilinear paths (wires).
Minimizing the cost of the network (which means minimizing the
total wire length) is desirable in terms of energy
consumption and signal interference. The additional requirement that
the terminal pairs are connected by \emph{shortest} rectilinear paths
aims at decreasing the interconnection delay (see Cong et
al.~\cite{clz-pdid-93} for a discussion in the context of rectilinear
Steiner arborescences, which have the same additional requirement; see
definition below).  Manhattan networks also arise in the area of
geometric spanner networks. Specifically, a minimum Manhattan network
can be thought of as the cheapest spanner under the $L_1$-norm for
given a set of points (allowing Steiner points). Spanners, in turn,
have numerous applications in network design, distributed algorithms,
and approximation algorithms.

MMN requires a Manhattan path between every terminal pair.  This
assumption is, however, not always reasonable.  Specifically in VLSI
design a wire connection is necessary only for a, often comparatively
small, subset of terminal pairs, which may allow for substantially
cheaper circuit layouts.  In this scenario, GMMN appears to be a more
realistic model than MMN.

\paragraph{Previous Work.}

MMN was introduced by Gudmundsson et al.~\cite{gln-ammn-01} who
gave 4- and 8-approximation algorithms for 2D-MMN
running in $O(n^3)$ and $O(n \log n)$ time, respectively.
The currently best known approximation algorithms for 2D-MMN have
ratio~2; they were obtained independently by Chepoi et
al.~\cite{cnv-raamm-08} using an LP-based method, by
Nouioua~\cite{n-eprmc-05} using a primal-dual scheme, and by Guo et
al.~\cite{gsz-gc2am-11} using a greedy approach.  The complexity of
2D-MMN was settled only recently by Chin et al.~\cite{cgs-mmnnp-11};
they proved the problem NP-hard.  It is not known whether 2D-MMN is
APX-hard. 

Less is known about MMN in dimensions greater than 2. Mu{\~n}oz et
al.~\cite{msu-mmnp3-09} proved that 3D-MMN is NP-hard to approximate
within a factor of 1.00002.  They also gave a constant-factor
approximation algorithm for a, rather restricted, special case 
of 3D-MMN.  More recently, Das et al. described the first
approximation algorithm for MMN in arbitrary, fixed dimension with a
ratio of~$O(n^\eps)$ for any $\eps > 0$ \cite{dgkksw-ammnh-11}.

GMMN was defined by Chepoi et al.~\cite{cnv-raamm-08} who asked
whether 2D-GMMN admits an $O(1)$-approximation. %
Apart from the formulation of this open problem, only special cases of
GMMN (such as MMN) have been considered %
so far.

One such special case (other than MMN) that has received significant
attention in the past is the \emph{rectilinear Steiner arborescence
  problem} (RSA).  Here, we are given $n$ terminals lying in the first
quadrant and the goal is to find a minimum-length rectilinear network 
that M-connects every terminal to the origin~$o$.
Hence, RSA is the special case of GMMN where~$o$ is
considered a (new) terminal and %
the set of terminal pairs
contains, for each terminal $t \ne o$, only the pair~$(o,t)$.  RSA was
introduced by Nastansky et al.~\cite{nss-cmtda-74} and has mainly been
studied in~2D.
2D-RSA is NP-hard~\cite{ss-rsapn-00}.
Rao et al.~\cite{rshs-rsap-92} gave a 2-approximation algorithm for
2D-RSA.  They also provided a conceptually simpler $O(\log
n)$-approximation algorithm based on rectilinear Steiner trees.  That
algorithm generalizes quite easily to dimensions $d > 2$ (as we show
in the appendix).
Lu et al.~\cite{lr-ptasrsap-00} and, independently,
Zachariasen~\cite{z-arsap-00} described polynomial-time approximation
schemes (PTAS) for RSA, both based on Arora's technique~\cite{a-asnph-03}.
Zachariasen pointed out that his PTAS can be generalized to the 
all-quadrant version of RSA but that it seems difficult to extend the
approach to higher dimensions.

\paragraph{Our Contribution.} 

Our main result is an $O(\log^{d+1} n)$-approximation algorithm for
GMMN (and, hence, MMN) in $d$ dimensions.
For the sake of simplicity, we first present our approach in 2D (see
Section~\ref{sec:polyl-appr-two}) and
then show how it can be generalized to higher dimensions (see
Section~\ref{sec:gener-high-dimens}).  We also provide an improved and
technically more involved $O(\log n)$-appro\-ximation for the special
case of 2D-GMMN, but this approach does not seem to generalize to
higher dimensions; see Section~\ref{sec:impr-algor-two}.  
To the best of our knowledge, we present the first approximation
algorithms for GMMN.  Our result for 2D is not quite the
constant-factor approximation that Chepoi et al.\ were asking for, but
it is a considerable step into that direction.  Note that the
poly-logarithmic ratio of our algorithm for GMMN in $d \ge 3$
dimensions constitutes an exponential improvement upon the previously  
only known approximation algorithm, which solves the special case
MMN, with a ratio of $O(n^\eps)$ for any $\eps>0$
\cite{dgkksw-ammnh-11}.

Our algorithm for GMMN is based on divide and conquer.  We identify
each terminal pair with its $d$-dimensional bounding box.
Consequently, we consider~$R$ a set of $d$-dimensional boxes.  We use
$(d-1)$-dimensional hyperplanes to partition~$R$ recursively into
sub-instances.  The base case of our partition scheme consists of GMMN
instances where all boxes contain a common point.  We solve the
resulting special case of GMMN by reducing it to RSA.  We have
postponed the running-time analysis to Appendix~\ref{sec:runningtime}.

\section{Polylogarithmic Approximation for Two Dimensions}
\label{sec:polyl-appr-two}

In this section, we present an $O(\log^2n)$-approximation algorithm
for 2D-GMMN.  The algorithm consists of a \emph{main algorithm} that
recursively subdivides the input instance into instances of so-called
\emph{$x$-separated} GMMN; see Section~\ref{sec:main-algorithm}.  We
prove that the instances of $x$-separated GMMN can be solved
independently by paying a factor of $O(\log n)$ in the approximation
ratio.  Then we show how to approximate $x$-separated GMMN within
ratio $O(\log n)$; see Section~\ref{sec:appr-x-separ}.  This yields an
overall ratio of $O(\log^2n)$.  

\subsection{Main Algorithm}
\label{sec:main-algorithm}

Our approximation algorithm is based on divide and conquer.  Let $R$
be the set of terminal pairs that are to be M-connected.  We identify
each terminal pair with its bounding box, that is, the smallest
axis-aligned rectangle that contains both terminals.  As a consequence
of this, we consider $R$ a set of rectangles.  Let $\xm$ be the
median in the multiset of the $x$-coordinates of terminals.  We
identify $\xm$ with the vertical line at $x=\xm$.

Our algorithm divides $R$ into three subsets $\Rl,\Rm$, and $\Rr$.
The set~$\Rl$ consists 
of all rectangles that lie {\em completely} to the left of the
vertical line $\xm$.  Similarly, the set $\Rr$ consists of all
rectangle that lie {\em completely} to the right of~$\xm$.  The
set $\Rm$ consists of all rectangles that intersect $\xm$.

We consider the sets \Rl, \Rm, and \Rr as separate instances
of GMMN and apply 
our algorithm recursively to~\Rl and to~\Rr.  The union of the two
resulting networks is a rectilinear network that M-connects all
terminal pairs in~$\Rl\cup\Rr$. %

It remains to M-connect the pairs in~\Rm.  We call an GMMN instance
(such as~\Rm) \emph{$x$-separated} if there is a vertical line (in our
case $\xm$) that intersects every rectangle.  We exploit this property
to design a simple $O(\log n)$-approximation algorithm for
$x$-separated GMMN; see Section~\ref{sec:appr-x-separ}.  Later, in
Section~\ref{sec:impr-algor-two}, we improve upon this and describe an
$O(1)$-approximation algorithm for $x$-separated GMMN.

To analyze the performance of our main algorithm, let $\rho(n)$ denote
the algorithm's worst-case 
approximation ratio for instances with $n$ terminal
pairs.  Now assume that our input instance $R$ is a worst case.  More
precisely, the cost of the solution of our algorithm \emph{equals}
$\rho(n)\cdot\opt$, where $\opt$ denotes the cost of an optimum
solution~\Nopt to~$R$.  Let~\Nl and~\Nr be the parts of~\Nopt to the
left and to the right of~\xm, respectively.

Due to the choice of $\xm$, at most $n$ terminals lie to the left of
$\xm$.  Therefore, $\Rl$ contains at most $n/2$ terminal pairs.  Since
$\Nl$ is a feasible solution to $\Rl$, we conclude that the cost of
the solution to $\Rl$ computed by our algorithm is bounded by
$\rho(n/2)\cdot\|\Nl\|$, where $\| \cdot \|$ measures the length of a
network.  Analogously, the cost of the solution 
computed for $\Rr$ is bounded by $\rho(n/2)\cdot\|\Nr\|$.  Now we assume that we
can approximate $x$-separated instances with a ratio of
$\rho_x(n)$.  Since $\Nopt$ is also a feasible solution to
the $x$-separated instance $\Rm$, we can compute a solution of cost
$\rho_x(n)\cdot\opt$ for $\Rm$.

Therefore, we can bound the total cost of our algorithm's solution $N$ to~$R$ by
\begin{displaymath}
  \rho(n)\cdot\opt=\|N\|\leq \rho(n/2)\cdot(\|\Nl\|+\|\Nr\|)+\rho_x(n)\cdot\opt\,.
\end{displaymath}
Note that this inequality does not necessarily hold if $R$ is \emph{not} a worst case since then $\rho(n)\cdot\opt>\|N\|$.  The networks~$\Nl$ and~$\Nr$ are separated by~$\xm$, hence they are
edge disjoint and $\|\Nl\|+\|\Nr\|\leq\opt$.  This yields the
recurrence $\rho(n)\leq \rho(n/2)+\rho_x(n)$, which resolves to
$\rho(n)=\log n \cdot\rho_x(n)$.  Let's summarize this discussion. 

\begin{lemma}
  \label{lem:approx-2d}
  If $x$-separated 2D-GMMN admits a $\rho_x(n)$-approximation, 
  2D-GMMN admits a $(\rho_x(n)\cdot\log n)$-approximation.
\end{lemma}

Combining this lemma with our $O(\log n)$ approximation algorithm for
$x$-separated instances described below, we obtain the following
intermediate result.

\begin{theorem}
  2D-GMMN admits an $O(\log^2 n)$-approximation.
\end{theorem}

\subsection{Approximating $x$-Separated Instances}
\label{sec:appr-x-separ}

In this section, we describe a simple algorithm for approximating
$x$-separated 2D-GMMN instances with a ratio of $O(\log n)$.  Let~$R$
be our input. %
Since~$R$ is $x$-separated, all rectangles in~$R$ intersect a common
vertical line.  W.l.o.g., this is the $y$-axis.

The algorithm works as follows.  Analogously to the main algorithm
presented in Section~\ref{sec:main-algorithm}, we recursively
subdivide the $x$-separated input instance, but this time according to
the \emph{$y$-coordinate}. %
As a result of this, the input instance
$R$ is decomposed into \emph{$y$-separated} sub-instances. Moreover,
since each of these sub-instances is (as a subset of $R$) already
$x$-separated, we call these instances \emph{$xy$-separated}.  In
Section~\ref{sec:appr-xy-separ}, we give a specialized algorithm for
$xy$-separated instances. %

Let $\rho_x(n)$ be the ratio of our algorithm for approximating
$x$-separated GMMN instances and let $\rho_{xy}(n)$ be the ratio of
our algorithm for approximating $xy$-separated GMMN instances.  In
Section~\ref{sec:appr-xy-separ}, we show that $\rho_{xy}(n)=O(1)$.
Then Lemma~\ref{lem:approx-2d} (by exchanging $x$- and
$y$-coordinates) implies that $\rho_x(n)=\log
n\cdot\rho_{xy}(n)=O(\log n)$.

\begin{lemma}
  $x$-separated 2D-GMMN admits an $O(\log n)$-approximation.
\end{lemma}

\subsection{Approximating $xy$-Separated Instances}
\label{sec:appr-xy-separ}

It remains to show that $xy$-separated GMMN instances can be
approximated within a constant ratio.  Let $R$ be such an instance.  We
assume, w.l.o.g., that it is the $x$- and the $y$-axis that intersect
all rectangles in~$R$, that is, all rectangles contain the
origin.  Let $\Nopt$ be an optimum solution to $R$.  Let 
$N$ be the union of $\Nopt$ with the projections of $\Nopt$ to the $x$-axis
and to the $y$-axis.  The total length of~$N$ is
$\|N\|\leq2\cdot\opt=O(\opt)$ since every line segment of~\Nopt is
projected either to the $x$-axis or to the $y$-axis but not to both.  The
crucial fact about $N$ is that this network contains, for every
terminal $t$ in $R$, an M-path from~$t$ to the \emph{origin}~$o$.
In other words, $N$ is a feasible solution to the RSA instance of
M-connecting every terminal in~$R$ to~$o$.  

\begin{wrapfigure}{r}{25ex}
  \includegraphics{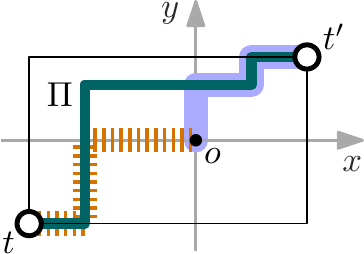}
  \caption{Network~$N$ connects all terminals to origin~$o$.}
  \label{fig:rsa+gmmn}
  \vspace*{-3ex}
\end{wrapfigure}
To see this, consider an arbitrary terminal pair $(t,t')\in R$.  Let
$\Pi$ be an M-path connecting~$t$ and~$t'$ in \Nopt; see
Fig.~\ref{fig:rsa+gmmn}.  Note that, since
the bounding box of $(t,t')$ contains~$o$, $\Pi$ intersects both $x$-
and $y$-axis.  To obtain an M-path from~$t$ to~$o$, we follow~$\Pi$
from $t$ to $t'$ until~$\Pi$ crosses one of the axes.  From that point
on, we follow the \emph{projection} of~$\Pi$ on this axis.  We
reach~$o$ when~$\Pi$ crosses the other axis; see the dotted path in
Fig.~\ref{fig:rsa+gmmn}.  Analogously, %
we obtain an M-path from~$t'$ to~$o$.

Let $T$ be the set of terminals in $R$.  We have shown above that
there is a feasible solution $N$ of cost $O(\opt)$ to the RSA instance
with terminal set~$T$.  There is a PTAS for RSA in two dimensions
\cite{z-arsap-00,lr-ptasrsap-00}.  Using this PTAS, we can efficiently
\emph{compute} a feasible RSA solution $N'$ for $T$ of cost
$O(1)\cdot\|N\|=O(\opt)$.  Moreover, $N'$ is also a feasible solution
to the GMMN instance~$R$.  To see this, note that~$N'$ contains, for
every terminal pair $(t,t') \in R$, an M-path~$\pi$ from $t$ to~$o$
and an M-path $\pi'$ from~$o$ to~$t'$.  Concatenating~$\pi$ and~$\pi'$
yields an M-path from~$t$ to~$t'$ as the bounding box of~$(t,t')$
contains~$o$. %
Thus we obtain the following result.

\begin{lemma}
  $xy$-separated 2D-GMMN admits a constant-factor approximation.
\end{lemma}

\section{Generalization to Higher Dimensions}
\label{sec:gener-high-dimens}

In this section, we describe an $O(\log^{d+1}n)$-approximation
algorithm for GMMN in $d$ dimensions, which is a
generalization of the algorithm for two dimensions presented in
Section~\ref{sec:polyl-appr-two}.  Let us view this algorithm from the
following perspective.  In Section~\ref{sec:main-algorithm}, we
reduced GMMN to solving $x$-separated sub-instances at the expense of a
$(\log n)$-factor in the approximation ratio (see
Lemma~\ref{lem:approx-2d}).  Applying the same lemma to the
$y$-coordinates in Section~\ref{sec:appr-x-separ}, we further reduced
the problem to solving $xy$-separated sub-instances, that is, to
instances that were separated with respect to both dimensions.  This
caused the second $(\log n)$-factor in our approximation ratio.
Finally, we were able to approximate these completely separated
sub-instances within constant ratio by solving a related RSA problem
(see Section~\ref{sec:appr-xy-separ}).

These ideas generalize to higher dimensions.  An instance $R$ of
$d$-dimensional GMMN is called \emph{$j$-separated} for some $j\leq d$
if there exist values $s_1,\dots,s_j$ such that, for each terminal pair
$(t,t')\in R$ and for each dimension $i\leq j$, we have that $s_i$
\emph{separates} the $i$-th coordinates $x_i(t)$ of $t$ and 
$x_i(t')$ of $t'$ (meaning that either $x_i(t)\leq s_i\leq x_i(t')$ or 
$x_i(t')\leq s_i\leq x_i(t)$).  Under this terminology, an
arbitrary instance of $d$-dimensional GMMN is always \emph{0-separated}.

We first show that if we can approximate $j$-separated GMMN with ratio
$\rho_j(n)$ then we can approximate $(j-1)$-separated GMMN with
ratio $\rho_j(n)\cdot\log n$; see Section~\ref{sec:separation}.  Then 
we show that $d$-separated GMMN can be approximated within a factor
$\rho_d(n)=O(\log n)$; see Section~\ref{sec:appr-d-sep}.  Combining
these two facts and applying 
them inductively to arbitrary (that is, 0-separated) GMMN instances
yields the following central result of our paper.

\begin{theorem}
  \label{thm:gener-high-dimens}
  GMMN in $d$ dimensions admits an $O(\log^{d+1}n)$-approximation.
\end{theorem}

As a byproduct of this algorithm, we obtain an
$O(\log^{d+1}n)$-approximation algorithm for \emph{MMN} where $n$
denotes the number of \emph{terminals}.  This holds since any MMN
instance with $n$ terminals can be considered an instance of GMMN
with $O(n^2)$ terminal pairs.  

\begin{corollary}
  \label{cor:mmn}
  \sloppy
  MMN in $d$ dimensions admits an $O(\log^{d+1} n)$-approximation,
  where $n$ denotes the number of terminals.
\end{corollary}

\subsection{Separation}
\label{sec:separation}

In this section, we show that if we can approximate $j$-separated GMMN
instances with ratio $\rho_j(n)$, we can approximate $(j-1)$-separated
instances with ratio $\log n\cdot\rho_j(n)$.  The separation algorithm
and its analysis work analogously to the main algorithm for 2D where
we reduced (approximating) %
2D-GMMN to (approximating) %
$x$-separated 2D-GMMN; see Section~\ref{sec:main-algorithm}

Let $R$ be a set of $(j-1)$-separated terminal pairs.  Let~\xm be the
median in the multiset of the $j$-th coordinates of terminals.  We
divide $R$ into three subsets~\Rl, \Rm, and \Rr.  The set~\Rl consists
of all terminal pairs $(t,t')$ such that $x_j(t),x_j(t')\leq\xm$ and
$\Rr$ contains all terminal pairs $(t,t')$ with
$x_j(t),x_j(t')\geq\xm$.  The set~$\Rm$ contains the remaining
terminal pairs, all of which are separated by the hyperplane
$x_j=\xm$.  We apply our algorithm recursively to~\Rl and~\Rr.  The
union of the resulting networks is a rectilinear network that
M-connects all terminal pairs $\Rl \cup \Rr$.

In order to M-connect the pairs in $\Rm$, we apply an approximation
algorithm for $j$-separated GMMN of ratio $\rho_j(n)$.  Note that the
instance $\Rm$ is in fact $j$-separated by construction.  The analysis
of the resulting algorithm for $(j-1)$-separated GMMN is analogous to
the 2D-case (see Section~\ref{sec:main-algorithm}) and is therefore
omitted.

\begin{theorem} 
  Let $1\leq j\leq d$.  If $j$-separated GMMN admits a
  $\rho_j(n)$-approxi\-ma\-tion, then $(j-1)$-separated GMMN admits a
  $(\rho_j(n)\cdot\log n)$-approximation.
\end{theorem}

\subsection{Approximating $d$-Separated Instances}
\label{sec:appr-d-sep}

In this section, we show that we can approximate instances of
$d$-separated GMMN within a ratio of $O(\log n)$ by reducing the
problem to RSA.
Let $R$ be a $d$-separated instance and let $T$ be the set of all terminals in
$R$. %
As $R$ is $d$-separated,
all bounding boxes defined by terminal pairs in $R$ contain a common
point, which is, w.l.o.g., the origin.

As in the two-dimensional case (see Section~\ref{sec:appr-xy-separ}),
we M-connect all terminals to the origin by solving an RSA instance
with terminal set $T$. This yields a feasible GMMN solution to~$R$ since for each pair 
$(t,t') \in R$ there is an M-path from $t$ to the origin and an M-path from the origin to $t'$.  The union of these paths is an M-path from $t$ to $t'$ since the origin is contained in the bounding box of $(t,t')$.

 Rao et al.~\cite{rshs-rsap-92} presented an 
$O(\log |T|)$-approximation algorithm for 2D-RSA, which 
generalizes, in a straight-forward manner, to $d$-dimensional RSA;
see Appendix~\ref{sec:gen-rsa} for details. 
Hence, we can use the algorithm of Rao et al.\ to efficiently compute
a feasible GMMN solution.  The following lemma shows that this
solution is in fact an $O(\log n)$-approximation.  
The proof is similar to the proof of Lemma~7 in the paper of Das et
al.\ \cite{dgkksw-ammnh-11}. 

\begin{lemma}
  \label{lem:lgn-approx-gmmn}
  $d$-separated GMMN admits an $O(\log n)$-approximation for any fixed
  number~$d$ of dimensions.
\end{lemma}
\begin{pf}
  Below we show that there is a solution of cost $O(\opt)$ to the RSA
  instance connecting $T$ to the origin.  Observing that $|T|\leq 2n$
  and using our extension of the $O(\log |T|)$-approximation algorithm
  of Rao et al.\ (see 
  Appendix~\ref{sec:gen-rsa}), we can efficiently compute a
  feasible GMMN solution of cost $O(\opt \cdot \log n)$.

We now show that there is an RSA solution of cost $O(\opt)$.  Let $\Nopt$ be an optimal GMMN solution to $R$ and let
$N$ be the projection of $\Nopt$ onto all subspaces that are spanned by some subset of the coordinate axes. Since there are $2^d$ such subspaces, which is a constant for fixed $d$, the cost of $N$ is $O(\opt)$.

It remains to show that $N$ M-connects all terminals to the origin,
that is, $N$ is a feasible solution to the RSA instance. 
First, note that $\Nopt\subseteq N$ since we project on the $d$-dimensional space, too.  Now consider an arbitrary terminal pair $(t,t')$ in~$R$
and an M-path $\pi$ in $\Nopt$ that M-connects $t$ and $t'$. Starting
at~$t$, we
follow~$\pi$ until we reach the first point~$p_1$ where one of the 
coordinates becomes zero.  W.l.o.g., $x_1(p_1)=0$.  Clearly~$\pi$
contains such a point as the bounding box of~$(t,t')$ contains the origin.  
We observe that $p_1$ lies in the 
subspace spanned by the $d-1$ coordinate axes $x_2,\dots,x_d$.
From $p_1$ on we follow the projection of $\pi$ onto this
subspace until we reach the first point $p_2$ where another coordinate
becomes zero; w.l.o.g., $x_2(p_2)=0$. 
Hence, $p_2$ has at least two coordinates that are zero, that is, $p_2$
lies in a subspace spanned by only $d-2$ coordinate axes.
Iteratively, we continue following the projections of~$\pi$ onto
subspaces with a decreasing number of dimensions until every coordinate is
zero, that is, we have reached the origin. An analogous argument shows that
$N$ also contains an M-path from $t'$ to the origin. 
\end{pf}

\section{Improved Algorithm for Two Dimensions}
\label{sec:impr-algor-two}

In this section, we show that 2D-GMMN admits an $O(\log
n)$-approximation, which improves upon the $O(\log^2 n)$-result of
Section~\ref{sec:polyl-appr-two}.  To this end, we develop a
$(6+\eps)$-approximation algorithm for $x$-separated 2D-GMMN, for any
$\eps>0$.  While the algorithm is simple, its analysis turns out to be
quite intricate.  In Appendix~\ref{sec:tightness}, we show tightness. 
Using Lemma~\ref{lem:approx-2d}, our new subroutine
for the $x$-separated case yields the following.

\begin{theorem}
  2D-GMMN admits a $((6+\eps)\cdot\log n)$-approximation.
\end{theorem}

Let $R$ be the set of terminal pairs of an $x$-separated instance of
2D-GMMN.  We assume, w.l.o.g., that each terminal pair
$(l,r)\in R$ is separated by the $y$-axis, that is, $x(l) < 0 \le x(r)$.
Let \Nopt be an optimum solution to~$R$.  Let \optver and \opthor be the
total costs of the vertical and horizontal segments in \Nopt,
respectively.  Hence, $\opt = \optver+\opthor$.  We first
compute a set~$S$ of horizontal line segments of total cost $O(\opthor)$
such that each rectangle in~$R$ is \emph{stabbed} by some line
segment in~$S$; see Sections~\ref{sec:stabbing-right-part}
and~\ref{sss:both}.  Then we M-connect the terminals to the $y$-axis
so that the resulting network, along with the affected part of the
$y$-axis and the stabbing~$S$, forms a feasible solution to~$R$ of cost
$O(\opt)$; see Section~\ref{sec:algorithm}.

\subsection{Stabbing the Right Part}
\label{sec:stabbing-right-part}

We say that a horizontal line segment~$h$ {\em stabs} an axis-aligned
rectangle~$r$ if $h$ intersects the boundary of~$r$ twice.  A set %
of horizontal line segments is a \emph{stabbing} of a set of
axis-aligned rectangles if each rectangle is stabbed by some line
segment. %
For any geometric object, let its \emph{right part} be its
intersection with the closed half plane to the right of the $y$-axis.
For a \emph{set} of objects, let its right part be the set of the
right parts of the objects.
Let~\Bright be the right part of~$R$,
let~\Nright be the right part of~\Nopt,
and let~\Nrighthor be the set of horizontal line segments in~\Nright.
In this section, we show how to construct a stabbing of~\Bright of
cost at most $2\cdot\|\Nrighthor\|$.

For $x' \ge 0$, let $\ell_{x'}$ be the vertical line at $x=x'$.  Our algorithm
performs a left-to-right sweep starting with~$\ell_0$.  Note that,
for every $x\geq 0$, the intersection of~\Bright with~$\ell_x$ forms a
set $\I_x$ of intervals.  The intersection of $\Nrighthor$ ($\bigcup
\Nrighthor$ to be precise) with~$\ell_x$
is, at any time, a set of points that constitutes %
a \emph{piercing} for~$\I_x$, that is,
every interval in $\I_x$ contains a point in $\ell_x\cap
\Nrighthor$.  Note that $\|\Nrighthor\|=\int |\ell_x\cap \Nrighthor|dx$.

We imagine that we continuously move~$\ell_x$ from $x=0$ to the right.
At any time, we maintain an inclusion-wise minimal piercing~$P_x$
of~$\I_x$.  With increasing~$x$, we only \emph{remove} points
from~$P_x$; we never add points.  This ensures that the traces of the
points in~$P_x$ form horizontal line segments that all touch the
$y$-axis.  These line segments form our stabbing of~\Bright.

The algorithm proceeds as follows.  It starts at $x:=0$ with an
arbitrary minimal piercing~$P_0$ of~$\I_0$.  Note that we can even
compute an optimum piercing~$P_0$.  We must adapt~$P_x$
whenever~$\I_x$ changes.  With increasing~$x$, $\I_x$ decreases
inclusion-wise since all rectangles in~\Bright touch the $y$-axis.  So
it suffices to adapt the piercing~$P_x$ at \emph{event points}; $x$ is
an event point if and only %
if $x$ is the $x$-coordinate of a right edge of a rectangle in~$R^+$.

Let $x'$ and $x''$ be consecutive event points.  Let $x$ be such that
$x' < x \le x''$.  Note that $P_{x'}$ is a piercing for~$\I_x$ since
$\I_x \subset \I_{x'}$.  The piercing~$P_{x'}$ is, however, not
necessarily \emph{minimal} w.r.t.~$\I_x$.  When the sweep line
passes~$x'$, we therefore have to drop some of the points in~$P_{x'}$
in order to obtain a new minimal piercing.  This can be done by
iteratively removing points from~$P_{x'}$ such that the resulting set
still pierces~$\I_x$.  We stop at the last event point
(afterwards, $\I_x=\emptyset$) and output the traces of the piercing.

\begin{figure}[tb]
  \subfloat[$S^+$ stabs~$R^+$, and $S$ stabs~$R$.\label{sfg:stabbing-both}]%
  {\includegraphics{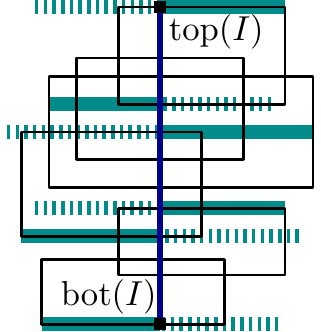}}
  \hfill
  \subfloat[$N=\rsaUp \cup \rsaDown \cup S$ is feasible
  for~$R$.\label{sfg:nisfeasible}]%
  {\includegraphics{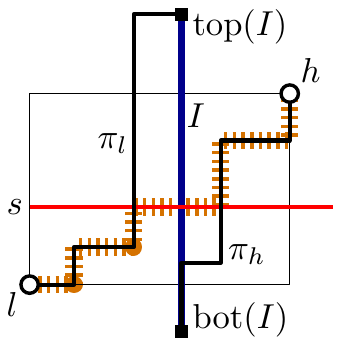}}
  \hfill
  \subfloat[$\Nopt \cup \{I\}$ is feasible for RSA instances
  $(L,\topp)$, $(H,\bott)$.\label{sfg:i+noptisrsa}]%
  {\hspace{8.8ex}\includegraphics{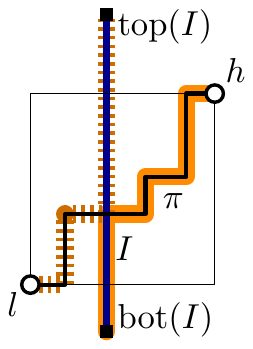}\hspace{8.8ex}}
  \caption{The improved algorithm for $x$-separated 2D-GMMN.}
  \label{fig:x-separated}
\end{figure}

It is clear that the algorithm produces a stabbing of~\Bright; see the
thick solid line segments in Fig.~\ref{sfg:stabbing-both}.  The
following lemma is crucial to prove the overall cost of the stabbing.

\begin{lemma}
  For any $x\geq 0$, it holds that $|P_x|\leq 2\cdot |\ell_x\cap \Nrighthor|$.
\end{lemma}

\begin{pf}
  Since $P_x$ is a minimal piercing, there exists, for every $p\in
  P_x$, a \emph{witness interval} $I_p\in \I_x$ that is pierced by $p$
  but not by $P_x \setminus \{p\}$.  Otherwise we could remove~$p$
  from~$P_x$, contradicting the minimality of~$P_x$.

  Now we show that an arbitrary point~$q$ on~$\ell_x$ is contained in the
  witness intervals of at most two points in~$P_x$.  Assume, for the
  sake of contradiction, that~$q$ is contained in the witness
  intervals of points $p, p', p'' \in P_x$ with strictly increasing
  $y$-coordinates.  Suppose that~$q$ lies above~$p'$.  But then the
  witness interval~$I_p$ of~$p$, which contains~$p$ and~$q$, must
  also contain~$p'$, contradicting the definition of~$I_p$.  The
  case~$q$ below~$p'$ is symmetric.

  Recall that $\ell_x\cap \Nrighthor$ is a piercing of $\I_x$ and, hence,
  of the $|P_x|$ many witness intervals.  Since every point in
  $\ell_x\cap \Nrighthor$ pierces at most two witness intervals, the
  lemma follows.
\end{pf}

Observe that the cost of the stabbing is $\int |P_x|dx$.  By the above
lemma, the cost of the stabbing can be bounded by
$\int |P_x|dx \leq \int 2 \cdot |\ell_x \cap \Nrighthor| dx =
2\cdot \|\Nrighthor\|$, which proves the following lemma.

\begin{lemma}
  \label{lem:stabhalf}\sloppy
  Given a set~$R$ of rectangles intersecting the $y$-axis, we can
  compute a set of horizontal line segments of cost at most
  \mbox{$2\cdot\opthor$} that stabs~\Bright. %
\end{lemma}

\subsection{Stabbing Both Parts}
\label{sss:both}

We now detail how we construct a stabbing of~$R$.  To this end we apply 
Lemma~\ref{lem:stabhalf} to
compute a stabbing~\Sleft of cost at most $2\cdot\|\Nlefthor\|$ for the
left part~\Bleft of~$R$ and a stabbing~\Sright of cost at most $2\cdot
\|\Nrighthor\|$ for the right part~\Bright.  Note that $\Sleft \cup
\Sright$ is not necessarily a stabbing for~$R$ since there can be
rectangles that are not \emph{completely} stabbed by one segment.  To
overcome this difficulty, we mirror~\Sleft and \Sright to the
respective other side of the $y$-axis; see
Fig.~\ref{sfg:stabbing-both}.  The total cost of the 
resulting set~$S$ of horizontal line segments is at most
$4(\|\Nlefthor\|+\|\Nrighthor\|) = 4\cdot\opthor$.  The set~$S$
stabs~$R$ since, for every rectangle~$r \in R$, the larger among its
two (left and right) parts is stabbed by some segment~$s$ 
and the smaller part is stabbed by the mirror image~$s'$ of~$s$.  Hence,
$r$ is stabbed by the line segment $s\cup s'$.
Let us summarize. 

\begin{lemma} 
  \label{lem:stabbing}\sloppy
  Given a set~$R$ of rectangles intersecting the $y$-axis, we can
  compute a set of horizontal line segments of cost at most
  \mbox{$4\cdot\opthor$} that stabs~$R$. %
\end{lemma}

\subsection{Connecting Terminals and Stabbing}
\label{sec:algorithm}

We assume that the union of the rectangles in~$R$ is connected.  Otherwise
we apply our algorithm separately to each subset of~$R$ that induces a
connected component of~$\bigcup R$.
Let $I$ be the line segment that is the intersection of the
$y$-axis with $\bigcup R$.  Let~\topp and~\bott be the top and bottom
endpoints of~$I$, respectively.  Let $L$ be the set containing every
terminal~$t$ with $(t,t') \in R$ and $y(t) \le y(t')$.
Symmetrically, let $H$ be the set containing every
terminal~$t$ with $(t,t') \in R$ and $y(t) > y(t')$.
Note that~$L$ and~$H$ are not necessarily disjoint.

Using a PTAS for
RSA~\cite{lr-ptasrsap-00,z-arsap-00}, we compute a near-optimal RSA
network~$\rsaUp$ connecting the terminals 
in~$L$ to~\topp and a near-optimal RSA network~$\rsaDown$ connecting the
terminals in~$H$ to~\bott. Then we return the network $N = 
\rsaUp \cup \rsaDown \cup S$,
where $S$ is the stabbing computed by the
algorithm in Section~\ref{sss:both}.

We now show that this network is a feasible solution to~$R$.
Let $(l,h) \in R$.  W.l.o.g., $l\in L$ and $h\in H$.  Hence, 
$\rsaUp$ contains a path~$\pi_l$ from~$l$ to~\topp, see
Fig.~\ref{sfg:nisfeasible}.  This path starts inside the 
rectangle~$(l,h)$ and leaves it through its top edge.  Before
leaving~$(l,h)$, the path intersects a line segment~$s$ in~$S$ that
stabs~$(l,h)$.  This line segment is also intersected by the
path~$\pi_h$ in~$\rsaDown$ that connects~$h$ to~\bott.
Hence, walking along~$\pi_l$, $s$,
and~$\pi_h$ brings us in a monotone fashion from~$l$ to~$h$. 

Now, let us analyze the cost of~$N$.  Clearly, the projection of~\Nopt
onto the $y$-axis yields the line segment~$I$.  Hence, $|I|
\le \optver$.  Observe that $\Nopt \cup \{I\}$ constitutes a solution
to the RSA instance $(L,\topp)$ connecting all terminals in~$L$ to~\topp and
to the RSA instance $(H,\bott)$ connecting all terminals in~$H$ to~\bott.
This holds since, for each terminal pair, its M-path~$\pi$ in~\Nopt
crosses the $y$-axis in~$I$; see Fig.~\ref{sfg:i+noptisrsa}.
Since~$\rsaUp$ and~$\rsaDown$ are near-optimal solutions to these RSA
instances, we obtain, for any $\eps >0$, that 
$\|\rsaUp\|\leq(1+\eps)\cdot\|\Nopt\cup I\|\leq (1+\eps)\cdot (\opt+\optver)$ 
and analogously $\|\rsaDown\|\leq (1+\eps)\cdot (\opt+\optver)$.

By Lemma~\ref{lem:stabbing}, we have that $\|S\| \le 4 \cdot \opthor$.
Assuming $\eps \leq 1$, this yields
\begin{align*}
  \|N\| & = \|\rsaUp\| + \|\rsaDown\| + \|S\|\\
  & \le (2+2\eps) \cdot (\opt+\optver) + 4\cdot \opthor\\
  & \le (2+2\eps)\cdot\opt+4\cdot(\optver+\opthor) \\
  & = (6 + \eps') \cdot \opt
\end{align*}
for $\eps' = \eps/2$, which we can make arbitrarily small by
making~\eps arbitrarily small.  We summarize our result as follows.

\begin{lemma}
  $x$-separated 2D-GMMN admits, for any $\eps>0$, a
  $(6+\eps)$-approximation.
\end{lemma}

\section{Conclusion and Open Problems}

We have presented an $O(\log^{d+1}n)$-approximation algorithm for
$d$-dimensional GMMN, which implies the same ratio for MMN.  Prior to
our work, no approximation algorithm for GMMN was known.  For $d \ge
3$, our result is a significant improvement over the ratio of
$O(n^{\eps})$ of the only approximation algorithm for
$d$-dimensional MMN known so far.

In 2D, there is still quite a large gap between the currently best
approximation ratios for MMN and GMMN.  Whereas we have presented an
$O(\log n)$-approximation algorithm for 2D-GMMN, 2D-MMN admits
2-approximations~\cite{cnv-raamm-08,gsz-gc2am-11,n-eprmc-05}---but is
2D-GMMN really harder to approximate than 2D-MMN?
Indeed, given that GMMN is more general than MMN, it may be possible
to derive stronger non-approximability results for GMMN.  So far, the only
such result is that 3D-MMN cannot be approximated beyond a factor of
1.00002~\cite{msu-mmnp3-09}.

Concerning the positive side, for $d \ge 3$, a constant-factor
approximation for $d$-dimen\-sional RSA would shave off a factor of
$O(\log n)$ from the current ratio for $d$-dimensional GMMN.  This may
be in reach given that 2D-RSA admits even a
PTAS~\cite{lr-ptasrsap-00,z-arsap-00}.  Alternatively, a
constant-factor approximation for $(d-k)$-separated GMMN for some
$k\le d$ would shave off a factor of $O(\log^k n)$ from the current
ratio for $d$-dimensional GMMN.

\section*{Acknowledgments}

We thank Michael Kaufmann for his hospitality and his enthusiasm
during our respective stays in T\"ubingen.  We thank Esther Arkin,
Alon Efrat, and Joe Mitchell for discussions.

\bibliographystyle{alpha}
\bibliography{../abbrv,../lncs,../ptas,../steiner,../spanner} 

\clearpage
\appendix

\noindent
\textbf{\large Appendix}

\section{Solving RSA in Higher Dimensions}
\label{sec:gen-rsa}

In this section, we show that we can approximate $d$-dimensional RSA
with a ratio 
of $O(\log n)$ even in the all-orthant case where every orthant may
contain terminals.  In this section, $n$ denotes the number of
\emph{terminals}.  We generalize the algorithm of Rao et 
al.\ \cite{rshs-rsap-92} who give an $O(\log n)$-approximation
algorithm for the one-quadrant version of 2D-RSA.

It is not hard to verify that the $O(\log n)$-approximation algorithm
of Rao et al.\ carries over to higher dimensions in a straightforward
manner if
all terminals lie in the \emph{same} orthant.  We can therefore obtain
a feasible solution to the all-orthant version by applying the
approximation algorithm to each orthant separately.  This worsens the
approximation ratio by a factor no larger than $2^d$ since there are
$2^d$
orthants. %
Hence, we can quite easily give an $O(\log n)$-approximation algorithm
for the all-orthant version since $2^d$ is a constant for fixed
dimension~$d$.

In what follows, we present a tailor-made approximation algorithm
for the all-orthant version of $d$-dimensional RSA that avoids the additional
factor of $2^d$.  Our algorithm is an adaption of the algorithm of Rao
et al., and our presentation closely follows their lines, too.

Consider an instance of RSA given by a set $T$ of terminals in $\R^d$
(without restriction of the orthant).  Let $o$ denote the origin.  The
algorithm relies on the following lemma, which we prove below.

\begin{lemma}
  \label{lem:raoetal}
  Given a rectilinear Steiner tree $B$ for terminal set $T\cup\{o\}$,
  we can find an RSA network $A$ for $T$ of length at most $\lceil
  \log_2 n \rceil\cdot\|B\|$.
\end{lemma}  

Every RSA network is also a rectilinear Steiner tree. Since the
rectilinear Steiner tree problem (RST) admits 
a PTAS for any fixed dimension $d$~\cite{a-ptase-JACM98},
we can generate a $(1+\eps)$-approximate RST network 
$B$ that connects $T$ and the origin. By means of
Lemma~\ref{lem:raoetal}, we get a $(1+\eps)\lceil\log_2
n\rceil$-approximation for the RSA instance~$T$.

\begin{theorem}
  \label{thm:kDimRSA}
  The all-orthant version of $d$-dimensional RSA admits a
  $(1+\eps) \cdot \lceil \log_2 n\rceil$ approximation for any
  $\eps > 0$.
\end{theorem}

Our proof of Lemma~\ref{lem:raoetal} relies on the following technical
lemma, which constitutes the main modification that we make to the
algorithm of Rao et al.  See Fig.~\ref{fig:min} for an illustration.

\begin{lemma}\label{lem:min-operation}
  Let $t,t'$ be two terminals.  Then we can compute in constant time a
  point $\min(t,t')$ and an M-path $\pi(t,t')$ from $t$ to $t'$
  containing $\min(t,t')$ with the following property.  The union of
  $\pi(t,t')$ with an M-path from $\min(t,t')$ to~$o$ M-connects %
  $t$ and~$t'$ to~$o$.
\end{lemma}

\begin{pf} %
  We start with the following simple observation.  If~$s$ and~$s'$ are
  points and~$p$ is a point in the bounding box $B(s,s')$ of~$s$
  and~$s'$, then M-connecting $s$ to $p$ and $p$ to $s'$ also
  M-connects $s$ to $s'$.

  Observe that the three bounding boxes $B(o,t)$, $B(o,t')$, and
  $B(t,t')$ have pairwise non-empty intersections.  By the Helly
  property of axis-parallel $d$-dimensional boxes, there exists a point
  $\min(t,t')$ that simultaneously lies in all three %
  boxes.

  M-connecting $\min(t,t')$ with~$t$ and~$t'$ yields $\pi(t,t')$, and
  M-connecting $\min(t,t')$ with~$o$ yields an M-path
  between any two of the points $t,t',o$ by a repeated application of
  the above observation.  This completes the proof.
\end{pf}

\begin{pf}[of Lemma~\ref{lem:raoetal}]
We double the edges of~$B$ and construct a Eulerian cycle~$C$ that 
traverses the terminals in $T\cup\{o\}$ in some order $t_0, t_1, \dots, t_n$.  
The length of~$C$ is at most~$2\|B\|$ by construction.  Now
consider the shortcut cycle $\tilde C$ in which we connect consecutive
terminals $t_i,t_{i+1}$ by the 
M-path $\pi(t_i,t_{i+1})$ as defined in Lemma~\ref{lem:min-operation}; we set
$t_{n+1}:=t_0$.  Clearly $\|\tilde C\|\leq\|C\|$.  We partition $\tilde
C$ it into two halves; $C_0 = \{ \pi(t_{2i}, t_{2i+1}) \mid 0 \le i \le
n/2 \}$ and $C_1 =\{\pi(t_{2i+1}, t_{2i+2}) \mid 0 \le i \le n/2-1\}$.
For at least one of the two halves, say~$C_0$, we have $\|C_0\| \le \|B\|$.

We use $C_0$ as a partial solution and recursively M-connect the
points in the set $T':=\{\min(t_{2i}, t_{2i+1}) \mid 0 \le i \le n/2\}$, which
lie in $C_0$ (see Lemma~\ref{lem:min-operation}), to 
the origin by an arboresence $A'$.  Lemma~\ref{lem:min-operation}
implies that the resulting network $A=C_0\cup A'$ is in fact a
feasible RSA solution.  The length of~$A$ is at most 
$\|C_0\| + \|A'\| \le \|B\| +\|A'\|$.
Note that $|T'|\leq (|T|+1)/2$.  

To summarize, we have described a procedure that, given the
rectilinear cycle~$C$ traversing terminal set $T\cup \{o\}$, computes
a shortcut cycle~$\tilde C$, its shorter half~$C_0$, and a new point
set~$T'$ that still has to be M-connected to the origin.  We refer to
this procedure as \emph{shortcutting}.  %

To compute the arboresence $A'$, observe that~$\tilde C$ is a
rectilinear cycle that traverses the points in~$T'$.  Shortcutting
yield a new cycle $\tilde C'$ of length at most
$\|\tilde C\|\leq\|C\|$, a half~$C_0'$ no longer
than~$\|B\|$, which we add to the RSA network, and a new point set
$T''$ of cardinality $|T''|\leq|T'|/2\leq (|T|+1)/4$, which we
recursively M-connect to the origin.

We repeat the shortcutting and recurse.  Each
iteration halves the number of new points, so the process
terminates in $O(\log n)$ iterations with a single point~$t$.  Since
$\min(o,p)=o$ for any point $p$ (see proof of
Lemma~\ref{lem:min-operation}) and our original terminal set
$T\cup\{o\}$ contained~$o$, we must have that $t=o$.  This shows that the
computed solution is feasible.  As each iteration adds length at most
$\|B\|$, we have $\|A\| \le \lceil \log_2 n \rceil \cdot \|B\|$.
\end{pf}

\section{Running Time Analysis} 
\label{sec:runningtime}

We first analyze the running times of the algorithm for $d>2$ in
Section~\ref{sec:gener-high-dimens}. 

Given an instance~$R$ of $0$-separated $d$-dimensional GMMN, the
algorithm uses $d$ recursive procedures to subdivide the problem into
$d$-separated instances.  For $j \in \{0,\dots,d-1\}$, let $T_j(n)$
denote the running time of the $j$-th recursive procedure.  The $j$-th
recursive procedure takes a $j$-separated instance $R$ as input and
partitions it into two $j$-separated instances, each of size at most
$|R|/2$, and one $(j+1)$-separated instance of size at most $|R|$.
The partitioning requires $O(n)$ steps for finding the median of the
$j$-th coordinate value of terminals in $R$. The two $j$-separated
instances are solved recursively and the $(j+1)$-separated instance is
solved with the $(j+1)$-th recursive procedure. Let $T_d(n)$ denote
the running time to solve a $d$-separated instance.  As pointed out in
Appendix~\ref{sec:gen-rsa}, we can approximate RSA in $d>2$ dimensions
by applying (an extension of) the algorithm of Rao et 
al.~\cite{rshs-rsap-92} to each orthant separately.  This requires
$O(n \log n)$ time as does the original algorithm.  Thus we have 
\begin{align*}
T_d(n) & = O(n \log n) & \\
T_{j}(n) & = 2T_{j}(n/2) + T_{j+1}(n) &\text{for }  j \in \{0,\dots,d-1\}
\end{align*}
The running time of our overall algorithm is given by $T_0(n)$. Solving
the recurrences above yields $T_0(n) = O(n\log^{d+1} n)$.   The
improved approximation algorithm of Theorem~\ref{thm:kDimRSA} uses
Aroras PTAS \cite{a-ptase-JACM98} for rectilinear Steiner trees, which
worsens the running time substantially but still leads to a polynomial
running time. 

Now we analyze the running time of the improved algorithm of
Section~\ref{sec:impr-algor-two}.  Stabbing $x$-separated instances
can be done with a sweep-line algorithm in $O(n \log n)$ time.  The
PTAS for RSA requires time $O(n^{1/\eps} \log n )$ for any \eps with 
$0 < \eps \le 1$.  Hence, we have that $T_1(n) =
O(n^{1/\eps} \log n)$, and $T_0(n) = 2T_0(n/2) + T_1(n)$.  Solving the
recursion yields a running time of $T_0(n) = O(n^{1/\eps}\log^2n)$ for the
improved algorithm.

\section{Example Showing the Tightness of Our Analysis}
\label{sec:tightness}

\begin{observation}
  There are infinitely many instances where the $O(\log n)$-approxi\-ma\-tion
  algorithm for 2D-GMMN described in Section~\ref{sec:impr-algor-two}
  has approximation performance $\Omega(\log n)$.
\end{observation}
\begin{pf}
  We recursively define an arrangement $A(n)$ of $n$ rectangles each
  of which represents a terminal pair; the lower left and upper right
  corner of the rectangle.  By $\alpha\cdot A(n)$ we denote the
  arrangement $A(n)$ but uniformly scaled in both dimensions so that
  it fits into an $\alpha\times\alpha$ square.  Let $\eps>0$ be a
  sufficiently small number.

  The arrangement $A(0)$ is empty.  The arrangement $A(n)$ consists of
  a unit square $S_n$ whose upper right vertex is the origin.  We add
  the arrangement $\Ar:=\eps\cdot A((n-1)/2)$ and place it in the
  first quadrant at distance $\eps$ to the origin. Finally, we add
  the arrangement $\Al:=(1-\eps)\cdot A((n-1)/2)$ inside the
  square $S_n$ so that it does not touch the boundary of $S_n$.  See
  Fig.~\ref{fig:tight2} for an illustration.

  \begin{figure}
    \begin{minipage}[b]{.48\textwidth}
      \centering
      \includegraphics{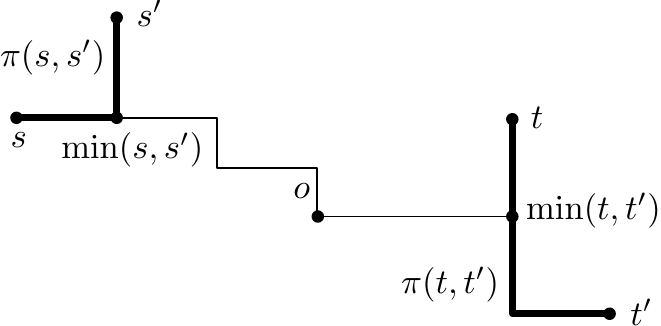}
      \caption{Illustration of Lemma~\ref{lem:min-operation}. For each
        terminal pair $t,t'$, we compute a suitable point $\min(t,t')$
        and an M-path $\pi(t,t')$ containing $\min(t,t')$.  Adding an
        arbitrary M-path from $\min(t,t')$ to $o$ M-connects $t$
        and $t'$ to~$o$.} 
      \label{fig:min}
    \end{minipage}
    \hfill
    \begin{minipage}[b]{.48\textwidth}
      \centering
      \includegraphics{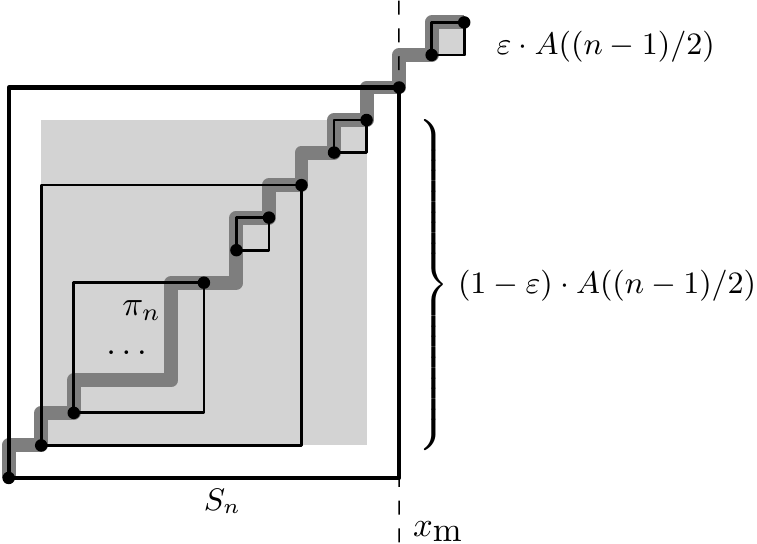}
      \caption{Recursive construction of the arrangement $A(n)$.  The
        gray M-path~$\pi_n$ shows an optimum solution.  The dashed
        vertical line marks where the algorithm separates.}
      \label{fig:tight2}
    \end{minipage}
  \end{figure}

  Let $\rho(n)$ denote the cost produced by our algorithm when applied
  to $A(n)$.  Observe that our algorithm partitions $A(n)$ into
  subinstances $\Rl=\Al$, $\Rm=\{S_n\}$, and $\Rr=\Ar$.  Solving the
  $x$-separated instance $\Rm$ by our stabbing subroutine costs~$1$.
  Let $\rho(n)$ be the cost of the solution to $A(n)$ that our
  algorithm computes. 
  Recursively solving $\Rl$ costs $(1-\eps)\cdot\rho((n-1)/2)$.
  Recursively solving $\Rr$ costs $\eps\cdot\rho((n-1)/2)$.
  Hence, the cost of the solution of our algorithm is $\rho(n)\geq
  1+\rho((n-1)/2)$.  %
  This resolves to $\rho(n)=\Omega(\log n)$.

  Finally, observe that the optimum solution is a single
  M-path~$\pi_n$ of length $1+2\eps$ going from the third to the first
  quadrant through the origin, see Fig.~\ref{fig:tight2}.
\end{pf}

\end{document}